\documentclass{elsart}
\usepackage{amssymb}
\usepackage{amsmath}\allowdisplaybreaks
\usepackage{graphicx}
\usepackage[normalem]{ulem}
\usepackage[dvips]{color}
\begin{document}
\begin{frontmatter}

\title{The amount of crustal entrainment and the type of Vela-like pulsars}

\author [XMU,KEY]{Ang Li\thanksref{info1}},
\author [Urumqi]{Jingbo Wang\thanksref{info2}},
\author [PKU]{Lijing Shao\thanksref{info3}},
\author [PKU]{and Ren-Xin Xu\thanksref{info4}}
\address[XMU]{Department of Astronomy and Institute of Theoretical Physics and Astrophysics, Xiamen University, Xiamen 361005, China} \address[KEY]{State Key Laboratory of Theoretical Physics, Institute of Theoretical Physics, Chinese Academy of Sciences, Beijing 100190, China}
\address[Urumqi]{Xinjiang Astronomical Observatory, Chinese Academy of Sciences, Urumqi, Xinjiang 830011, China}
\address[PKU]{School of Physics and State Key Laboratory of Nuclear Physics and Technology, Peking University, Beijing 100871, China}
\thanks[info1]{liang@xmu.edu.cn}
\thanks[info2]{wangjingbo@xao.ac.cn}
\thanks[info3]{lshao@pku.edu.cn}
\thanks[info4]{r.x.xu@pku.edu.cn}

\begin{abstract}
 The ``glitch crisis'' of Vela-like pulsars has been a great debate recently. It might challenge the standard two-component glitch model, because large fractions of superfluid neutrons are thought to be entrained in the lattices of the crust part, then there is not enough superfluid neutrons to trigger the large glitches in Vela-like pulsars. But the amount of entrainment which could effectively constrain the fractional moment of inertia of a pulsar, is very uncertain. In order to examine the importance of this parameter on the inner structures of neutron stars, we relax the ``glitch crisis'' argument, employ a set of most developed equations of state derived within microscopic many-body approaches that could fulfill the recent 2-solar-mass constraint, and evaluate their predictions for the fractional moment of inertia with two extremes of crustal entrainment. We find a final determination of the amount of entrainment could be closely related to the type of neutron star. If a large enough fraction of neutrons are entrained, a Vela-like pulsar could not be a hybrid star, namely no free quarks would be present in its core. In addition, we use the Vela data to narrow the parameter space of hyperon-meson couplings in the popular phenomenological relativistic mean field model.
\end{abstract}


\end{frontmatter}

\section{Introduction}           
\label{sect:intro}

Pulsars are usually regarded as highly-magnetized, rapidly rotating neutron stars (NSs), with periods between a few seconds and a millisecond. Pulsars emit a beam of electromagnetic radiation (most likely in radio or X-ray bands). The radiation from pulsars can be observed when the radiation beam is pointing toward the Earth, like a lighthouse. They are very stable rotators. A gradual spin-down is usually expected because of the loss of rotation energy via electromagnetic radiation and the acceleration of particles in their magnetospheres. A glitch is a sudden change of the pulsar rotation rate and one of the rotation irregularities which are often seen in young pulsars (age $< 10^5$ year).

There are two types of glitches whose physical mechanism, heavily depending on the inner structure of NSs, is still a matter of debate~\cite{Zhou14}. One type shows large glitches (e.g. the Vela pulsar; $\Delta \nu / \nu \sim 10^{-6}$) but releases negligible energy, while the other one, seen as large glitches in anomalous X-ray pulsars / soft gamma repeaters (AXPs/SGRs), is usually accompanied with detectable energy release, manifesting as X-ray bursts or outbursts. These observations
also urge a better understanding of the equation of state (EoS) of NSs.

Among them, a total of 17 glitches have been documented for the Vela pulsar (B0833-45) so far~\cite{Dodson02}, which makes it an ideal object for studying in detail the related physics mechanisms. Recently, its glitch activity parameter, defined in terms of the cumulative spin-up rate, has been used to constrain the underlying EoS of NS matter~\cite{Hooker13,Lattimer13,Link99}, based on a well-known two-component glitch model, in which one component refers to the charged component, and the other one is the superfluid component.

\begin{figure}
\centering
\includegraphics[width=9cm,angle=0]{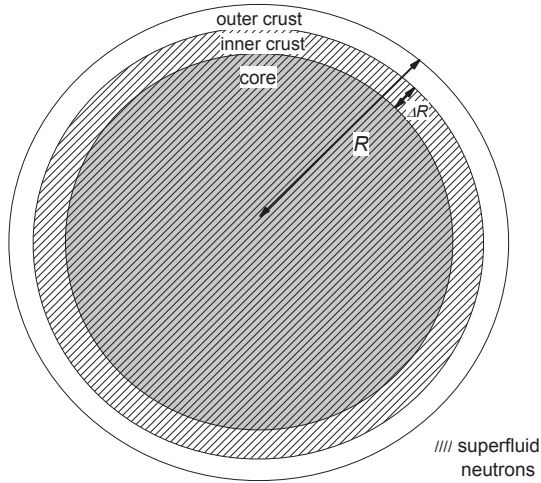}
\caption{Structure of a typical neutron star. The shaded area is where the superfluid neutrons are possible. The star radius is $R$ and $\Delta R$ is the thickness of the inner crust, defined as
the distance between neutron drip density and the core-crust
transition density.} \label{fig1}
\end{figure}

In the standard two-component glitch mechanism~\cite{Baym69}, the superfluid neutrons are limited in the inner crust of the star (shown in Fig.~\ref{fig1}), where the density has exceeded the neutron-drip density of $4.3\times 10^{11}$g$/$cm$^3$. It is a non-uniform system constituting neutron-rich nuclei, electrons and superfluid neutrons. Those neutron superfluid vortices are decoupled from the rotational evolution of the charged component of
the crust (and that part of the core that couples strongly
to it), since their interactions with the lattice of nuclei make them pinned to the crust. A glitch happens every time these two components recouple, when the lag between the two angular frequencies reaches a critical value, accompanied by a transfer of angular momentum from the superfluid component to the charged component.

``Glitch crisis'' was first proposed independently in Refs.~\cite{Chamel13,Andersson12}. It says that due to a large effective neutron mass (so-called \emph{crustal entrainment} effect by the crystal lattice) in the superfluid, there are not enough free superfluid neutrons to serve as the angular momentum reservoir of the star, then glitches observed in Vela-like pulsars can not be explained in the standard two-component glitch models, or more specifically, the vortex-mediated glitch interpretation. Several illuminating works, \cite{Hooker13,Link13,Piekarewicz14} are followed and have contributed a more comprehensive understanding of this problem. Also, this new glitch constraint, i.e., with crustal entrainment effect taken into account, has been used to select symmetry parameters of the nuclear interaction~\cite{Hooker13,Lattimer13}, which are both controversial and important to fundamental physics and astrophysical observational implications.

However, the argument of ``glitch crisis'' relies on the assumption  that the entire core of the star is coupled to the crust when glitches happen. This is suspicious because the glitch rise timescale ($<40$s) might be much smaller than the core-crust coupling time (about the order of days), which means only a fraction (maybe a half or less) of the core part could couple to the charged crust part and contribute to the total stellar moment of inertia~\cite{Link13}. Curial problems for determining the coupling timescale included at least the type of proton superconductivity in the core~\cite{Link13}, the details of how the vortices react to the pinning force, the realistic modeling of the pinning force in the crust~\cite{Pizzochero11,Haskell13}, and the impact of nuclear ``pasta'' phases near the crust bottom~\cite{Chamel13}. Presently they are very uncertain. In the present work we do not try to resolve these problems, but rather keep some reservations about the ``glitch crisis'' argument, and study two critical cases: the case with and without entrainment. The amount of crustal entrainment can be continuously varied between the two extremes of without entrainment and with entrainment.

We plan to investigate whether the amount of crustal entrainment is relevant for determining the amount of quark matter in the cores of Vela-like pulsars by making use of parameter-free microscopically-developed many-body approaches. In addition, the possibility of constraining hyperon parameters for phenomenological models from glitch observations will also be addressed, even that there is no problem for them to provide heavy enough hyperon stars~\cite{Weissenborn12,Vid11}. We leave the possibility of pure quark star (made entirely of deconfined $u, d, s$ quark matter) out of our picture, because it is usually regarded as bare and without a crust, therefore similar mechanism for triggering glitches in a NS would be of no meaning for a quark star. One should pursue a new explanation for the possibility that a Vela-like pulsar could be a quark star, which is out of scope of the present work.

The paper is organized as follows. In section 2, we describe our NS models with four microscopically-developed models in the market. In section 3, numerical results are discussed. A short summary is finally presented in section 4.

\section{Microscopic nuclear many-body models for NSs}
\label{sect:Mod}

As usual, we describe a NS as a spherically symmetric object in hydrostatic equilibrium, with gravity balanced by the pressure produced by the compressed nuclear matter. From the surface to its core, with the increase of density (as shown in Fig.~1), the following regions appear as follows:\\
 $1)~Outer~crust$ : Non-uniform coulomb lattice of neutron-rich nuclei embedded in a degenerate electron gas;\\
 $2)~Inner~crust$ : Non-uniform system of more exotic neutron-rich nuclei, degenerate electrons and superfluid neutrons;\\
 $3)~Core$ : Uniform nuclear matter in weak-interaction equilibrium (or $\beta-$equilibrium) with leptons. Strangeness is possible in the inner part of the core, such as hyperons (then a NS is called a hyperon star) and deconfined quarks (then a NS is called a hybrid star). One big puzzle is whether there is a certain amount of quark matter in the core of a NS. It has motivated many theoretical and experimental efforts in decades~\cite{Alford2005,liang08,Chen11}. As mentioned in the introduction, in this work, we try to connect it to the observations of pulsar glitches.

The employed microscopic models for nuclear matter EoS include the Brueckner-Hartree-Fock (BHF) approach~\cite{book}, a variational method~\cite{Var,Var2,APR}, the perturbative QCD model~\cite{Kurkela2010,Fraga14}, and the Dyson-Schwinger model~\cite{Chen11,Chen12,LiH11}. All the four methods have the ability to explain correctly the data obtained in both laboratorial heavy ion experiments and astrophysical observations.

The first two use only a realistic free nucleon-nucleon (NN) interaction as input, with parameters fitted to NN scattering phase shifts in different partial wave channels and to properties of the deuteron. The input realistic NN interaction adopted in the BHF approach is the Bonn B two-body force and a consistently-constructed three-body force (TBF)~\cite{LS08}. The inclusion of the microscopic TBF in the calculation for dense stellar matter and NS properties have been performed in many works~\cite{liang08,Chen11,book,Chen12,Schulze11,Schulze13,liang11,liang10,liang06,liang04}. The variational method EoS~\cite{Var,Var2} is the one calculated by Akmal, Pandharipande and Ravenhall (1998; hereafter APR). It used the Argonne $v_{18}$ two-body force and a Urbana three-body force. This APR EoS has been widely used and serves as a baseline EoS in recent articles~\cite{Kurkela2010}.

The latter two establish a link between nuclear matter properties and the underlying quantum chromodynamics (QCD) structure of strong interaction, although a first-principle calculation in such systems is unachievable due to the complicated nonlinear and nonperturbative nature of QCD. The present calculations of the perturbative QCD model~\cite{Kurkela2010,Fraga14} have been done for two massless and one massive quark flavors, with a quark chemical potential $\mu_B$ above $1$ GeV. The results are ready to be used for studying cold dense QCD matter, not only because it is shown to have a reasonably well convergence behaviour, but also the model has the ability to estimate its inherent systematic uncertainties via a dependence on the renormalization scale parameter. The Dyson-Schwinger model~\cite{Chen11,Chen12,LiH11} provides a continuum approach to QCD that can simultaneously address both confinement and dynamical chiral symmetry breaking. For the phenomenological extending of the effective interaction to finite chemical potential $\mu_B$, several studies on hybrid stars~\cite{Chen11,Chen12} have introduced a free parameter $\alpha$ to control the rate of approaching asymptotic freedom. A range of $\alpha=0\sim4$ has been used in their calculations.

\begin{figure}
\centering
\includegraphics[width=11cm,angle=0]{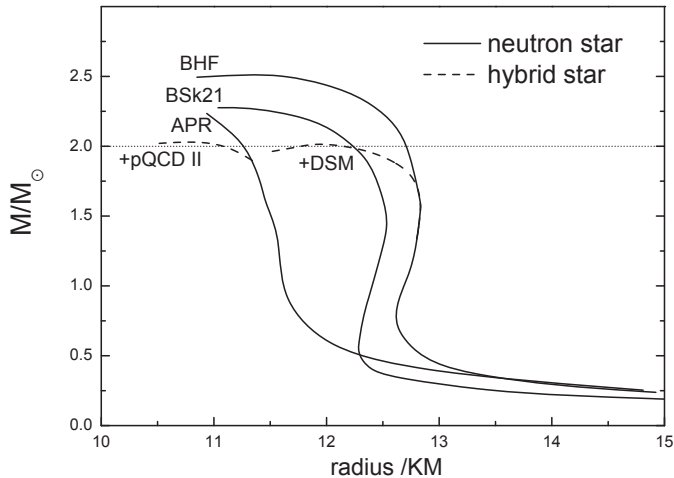}
\caption{Mass-radius relations of the NS models used in the present work. The BHF and ``+DSM'' data are taken from Chen et al. 2011, the APR and ``+pQCD II'' data are taken from Kurkela et al. 2010, and the BSk21 data is taken from Chamel et al. 2011. The recent 2-solar-mass constraint from the mass measurements of PSR J1614-2230 and PSR J0348+0432 is also shown.} \label{fig2}
\end{figure}

In the present work, for NSs, two EoS models are chosen: the variational method (labelled as APR) and the BHF approach (labelled as BHF). Two chosen models for hybrid stars are: the matching of an APR nuclear EoS with a perturbative QCD quark EoS (labelled as pQCD), and the matching of a BHF nuclear EoS with a Dyson-Schwinger quark EoS (labelled as DSM). The standard Gibbs construction is used for the matching. That is, the pressure $P$ is taken to be the same in the hadron-quark mixed phase to ensure mechanical stability, and would increase monotonically with baryon chemical potential $\mu_B$. In the mean time, a global charge neutrality is assumed. Under these matching criteria, there are two possibilities for the ``APR+pQCD'' matching, named as Case I and Case II by Kurkela et al. 2010. However, only Case II is considered in the present work for two reasons. First, Case I could only be possible for a large value of the renormalization scale; Secondly, the maximum mass of hybrid star in Case I is only $\sim1.75M_\odot$~\cite{Kurkela2010}, which is although comprisable but could be ruled out by the recent 2-solar-mass NS mass measurements~\cite{Demorest10,Antoniadis13}. We also notice that Case II happens for a high quark chemical potential, which means that the star is composed mostly of nucleonic matter and has only a small quark core.

Once the NS EoS $P(\mathcal{E})$, namely the pressure $P$ as a function of the energy density $\mathcal{E}$, is determined from a many-body approach, the mass-radius relation of a NS can be determined from the Tolman-Oppenheimer-Volkoff (TOV) equations,
\begin{equation}
 \frac{dP(r)}{dr}=-\frac{Gm(r)\mathcal{E}(r)}{r^{2}}
 \frac{\Big[1+\frac{P(r)}{\mathcal{E}(r)}\Big]
 \Big[1+\frac{4\pi r^{3}P(r)}{m(r)}\Big]}
 {1-\frac{2Gm(r)}{r}},
    \label{tov1:eps}
\end{equation}
\begin{equation}
\frac{dm(r)}{dr}=4\pi r^{2}\mathcal{E}(r),
    \label{tov2:eps}
\end{equation}
for the pressure $P$ and the enclosed mass $m(r)$ at a certain raidus $r$, where $G$ is the gravitational constant, and the light speed is assumed to be $c=1$.

The resulting star sequences employing various EoS models are collected in Fig.~2. All the chosen NS models can provide maximum masses as large as 2$M_\odot$, demanded by the recent astrophysical observations~\cite{Demorest10,Antoniadis13}. In the same figure, we also present the result of the BSk21 EoS, which has been used for the illustration of glitch crisis problem~\cite{Chamel13}. The BSk21 EoS~\cite{Chamel11} is the stiffest one among the three functionals, which are deduced from generalized Skyrme functionals simultaneously fitted to all the 2003 nuclear mass data and pure neutron matter calculations from microscopic models. The microscopic model for the BSk21 EoS is just the BHF approach~\cite{LS08} with Argonne $v_{18}$ two-body force and a consistently-constructed TBF mentioned above. One can see that the BSk21 case is slightly softer than the full BHF case. The NS maximum masses are 2.28$M_\odot$ and 2.51$M_\odot$, respectively. From the figure, one could expect that the predictions of the BSk21 EoS is in between the two cases of APR and BHF.

\section{Results}
\begin{figure}
\centering
\includegraphics[width=11cm,angle=0]{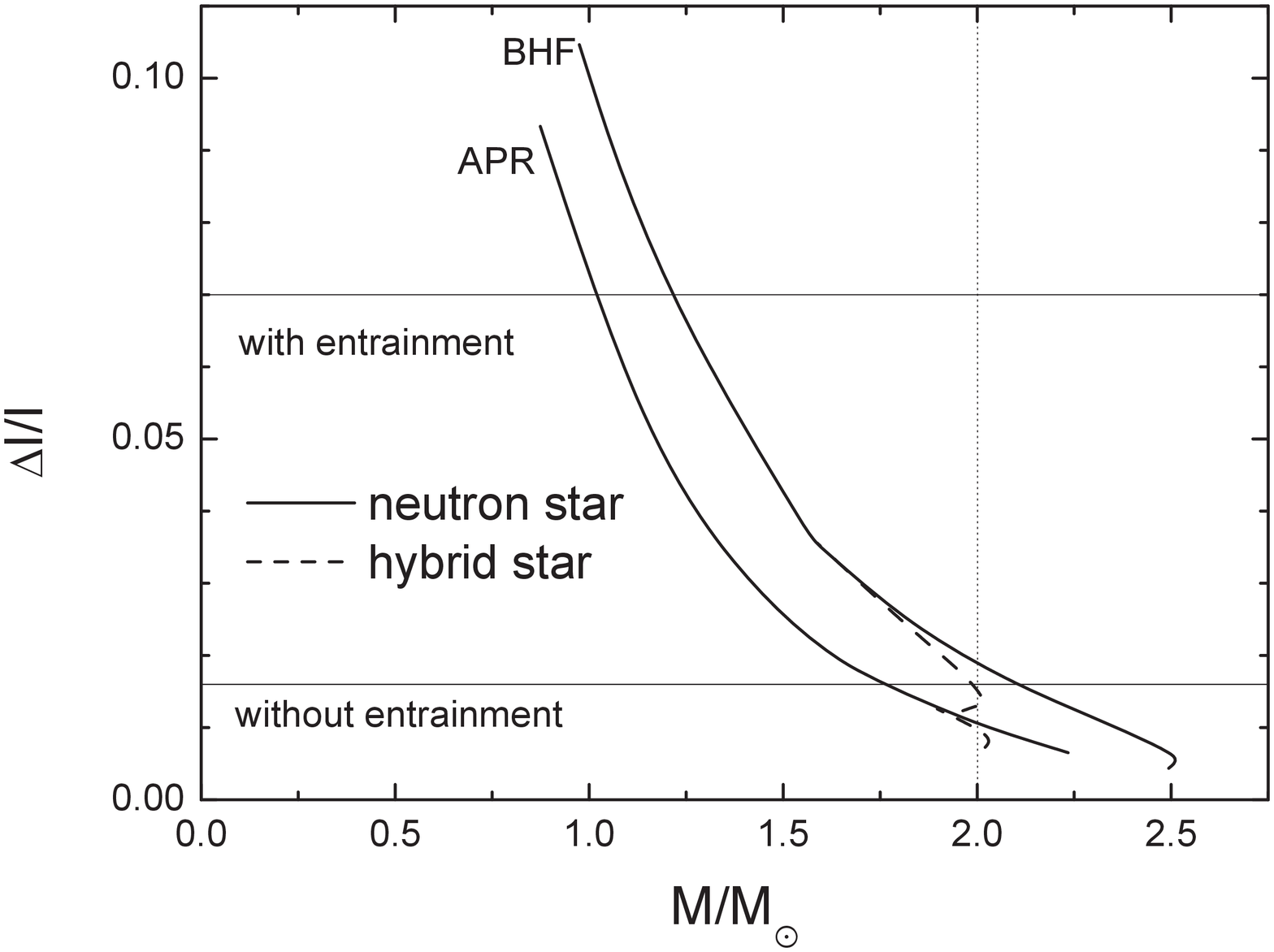}
\caption{Fractional moments of inertia for NSs and hybrid stars, as a function of stars' mass. The vertical line is the 2-solar-mass constraint from the mass measurements of PSR J1614-2230 and PSR J0348+0432.} \label{fig3}
\end{figure}
\subsection{Distinguish of NS models based on microscopic theories}

Using the subscript $s$ to refer to the superfluid component, and $c$ to the charged component (and that part of the core that couples strongly to it), from the Vela data, we have the following constraint for the crustal moment of inertia~\cite{Link99}:
\begin{equation}
   \frac{I_{s}}{I_{c}} = \frac{I_{\rm crust}-I_{\rm outer~crust}}{I-I_{s}} \simeq \frac{I_{\rm crust}}{I}\equiv \frac{\bigtriangleup I}{I} > \sim\! 0.016 \,.
 \label{VelaLimit3}
\end{equation}
with $I$ the total stellar moment of inertia (that which we observe). We have used the conditions that $I_{\rm outer~crust}\ll I_{s}$ and $I_{s}\ll I_{c}$. Based on this relation, reasonable limits were obtained for the Vela pulsar's radius with various of representative neutron star models~\cite{Link99}. That is $R\geq 12$km, assuming Vela's mass $M \sim 1.35 M_{\odot}$. While studies on the effective neutron mass in the superfluid indicate a much larger value than previous results, which severely limits the maximum number of superfluid neutrons that can possibly transfer angular momentum during glitches~\cite{Chamel12}. The ratio of neutrons to those free neutrons was estimated to be around $n/n_f = 4.3$~\cite{Andersson12}. With this taken into account, the pulsar constraint is modified to be $\bigtriangleup I/I > \sim\! 0.07$, and a nonphysically low mass limit would appear for a NS~\cite{Chamel13}, even with the quite stiff BSk21 EoS mentioned above. This is known as ``glitch crisis''.

The formula that they used for evaluating the fractional moment of inertia $\bigtriangleup I/I$ is updated in Ref.~\cite{Lattimer2007}:
\begin{equation}\label{di}
 \frac{\bigtriangleup I}{I}\simeq{8\pi p_tR^4\over3GM^2}\left[{MR^2\over I}-2\beta\right]e^{-4.8\Delta R/R}.
\end{equation}
with the compression parameter $\beta = GM/Rc^2$. We will use it for our present work. Essentially, the mass-radius relation $M(R)$, the core-crust transition pressure $p_t$, and the crust radius $\Delta R$ should be determined simultaneously and consistently from one basic theory which can describe both the low-density crust and the high-density core properly, however, no such theory is available presently. Theoretical and experimental uncertainties prevent them from being effective constraints~\cite{Lattimer13}.

Also, due to the facts that the crust contains a small fraction of the NS mass and usually much thinner than the core part, for our purpose, the core-crust transition pressure is fixed to be a middle value ($p_t=0.4{\rm~MeV~fm}^{-3}$) according to its favourable range: $0.2{\rm~MeV~fm}^{-3} \sim 0.65{\rm~MeV~fm}^{-3}$, constrained from experimental symmetry parameters~\cite{Lattimer2007}. We mention here that for the widely-used standard BPS crust EoS of Baym, Pethick, and Sutherland~\cite{BPS}, this value is $0.47{\rm~MeV~fm}^{-3}$. The fractional crust radius $\Delta R/R$ is also fixed to be $0.082$ since only a small deviation is expected away from this value for a typical $1.4 M_{\odot}$ star~\cite{Lattimer2007}. The resulting fractional moments of inertia are shown as a function of the star mass in Fig.~3.

From Fig.~3, we first notice that, in the case of with entrainment (when $\bigtriangleup I/I\geq0.07$ fulfilled), only very low mass NSs are allowed, with an upper mass limit in the small range of $1.05M_{\odot}\sim1.25M_{\odot}$ for three selected microscopic NS models (the plot of the Bsk21 case has been omitted for compactness), leaving a doubtful situation that whether they are still likely to be formed in a type II supernova explosion as generally believed. If the glitch constraint can be relaxed, to its lower limit of without-entrainment case (when $\bigtriangleup I/I \geq0.016$ fulfilled), the mass of a Vela-like pulsar should only be smaller than $1.77M_{\odot}\sim2.10M_{\odot}$, a constraint that can be easily fulfilled for a large population of pulsars based on many possible NS EoS models.

A proper value of this parameter should be in between, namely  $0.016<\bigtriangleup I/I<0.07$, although its determination largely relies on a more comprehensive understanding of the core-crust coupling. It is still possible to get many useful information from comparing the results of NSs and hybrid stars, for different
amount of crustal entrainment. The following discussions can be done based on the present models.

$1).$ If the inferred $\bigtriangleup I/I$ value is smaller than half of the value with entrainment (namely 0.07), namely in the range of $0.016\sim 0.035$, as commonly estimated in the literature~\cite{Lattimer13,Link13}, a Vela-like pulsar still has the chance to have a fraction of free quarks in its core, since it is supported by at least one of our microscopic models, the BHF model together with the DS model. However, it is not easy to distinguish a hybrid star with a normal NS with a same measured $\bigtriangleup I/I$, since their masses are very close to each other. It requires that the mass measurement of the star has at least an accuracy of $0.05\%$. That is, one can tell whether the star's mass is $1.997M_{\odot}$ or $2.102M_{\odot}$ if the fractional moment of inertia $\bigtriangleup I/I$ is measured to be 0.016. Presently, a mass measurement to an accuracy $0.05\%$ is plausible (see e.g. \cite{Weisberg10}) based on the post-Keplerian parameters from pulsar timing. The measurement of the moment of inertia from glitch observations is still model-dependent, while it is still challenging from pulsar timing~\cite{Damour88}.

$2).$ If the inferred $\bigtriangleup I/I$ value could be larger than the half of 0.07, namely a large enough fraction of neutrons are entrained, then it could not be a hybrid star, and should have a normal nuclear matter core, not a strange matter one. Notice that the possibility of stars with kaon condensation in its cores has been ruled out by several model calculations~\cite{liang08,Kurkela2010,liang10}.

\begin{figure}
\centering
\includegraphics[width=11cm,angle=0]{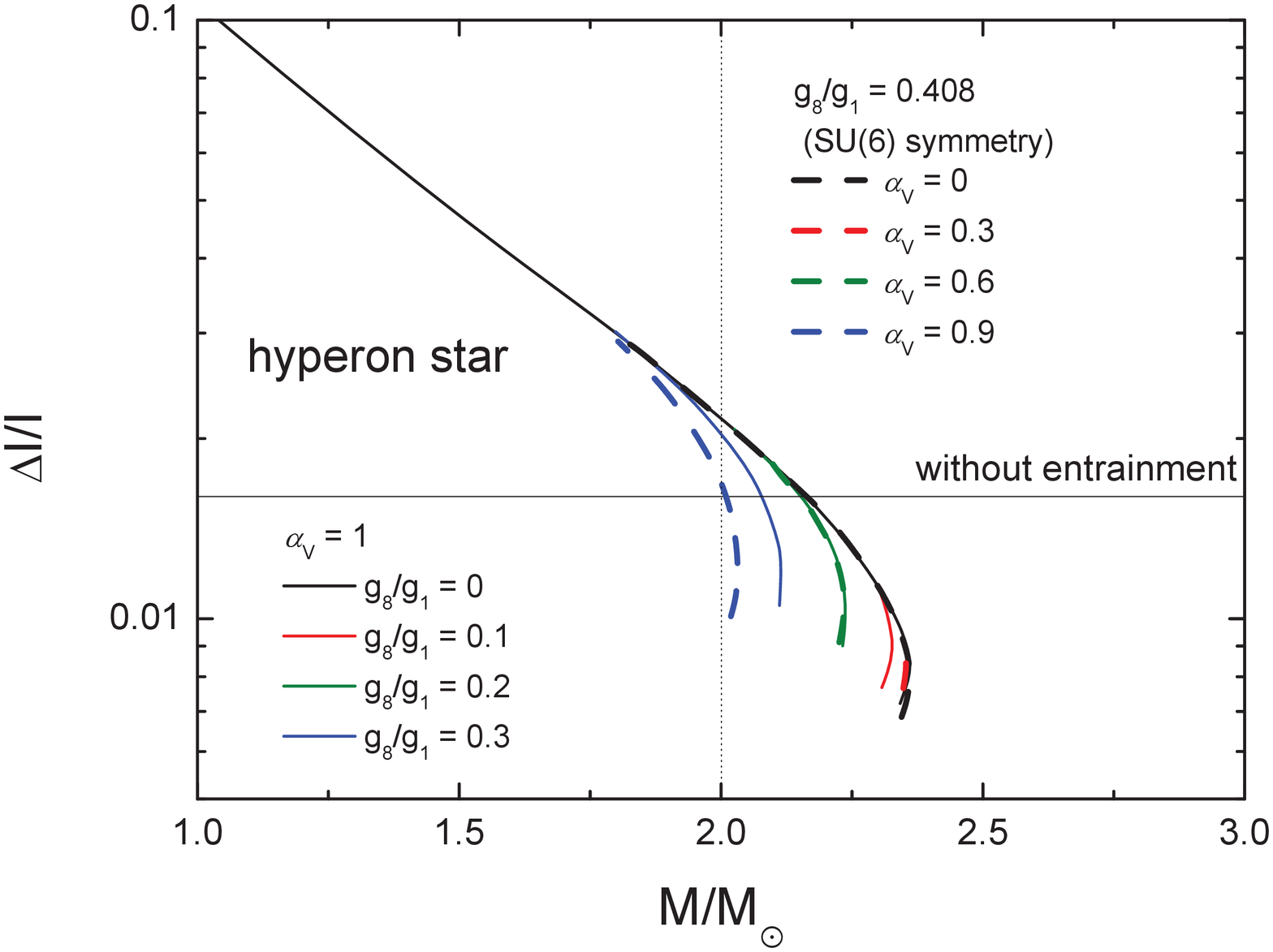}
\caption{Fractional moments of inertia for different sequences of hyperon stars as a function of stars' mass, with $g_8/g_1=0,0.1,0.2,0.3$ at a certain $\alpha_V=1$, and $\alpha_V=0,0.3,0,6,0.9$ at a certain $g_8/g_1=1/\sqrt{6}$. The vertical line is again the 2-solar-mass constraint from the mass measurements of PSR J1614-2230 and PSR J0348+0432.} \label{fig4}
\end{figure}
\subsection{Constraining hyperon coupling parameters of phenomenological models}

Although in the past years, the BHF approach has been extended in
order to include the hyperon degrees of freedom, and has been a powerful tool to explore properties of both hyperon stars and hypernuclei. However, the present poor knowledge of realistic hyperon potentials prevents a consistent result for hyperon star mass, with two recent convincing astrophysical observations. Specifically, the maximum mass is 1.37$M_\odot$~\cite{Schulze11,Schulze13,liang11} with the latest Nijmegen extended soft-core 08 hyperon-nucleon and hyperon-hyperon potentials, but two observations both reach 2$M_\odot$~\cite{Demorest10,Antoniadis13}. A way out may be the inclusion of improved hyperon-nucleon and hyperon-hyperon potentials and hyperonic three-body forces, which badly
needs further experimental data.

For the purpose of constraining hyperon parameter space based on the glitch data, we turn to one of the widely-used phenomenological relativistic mean field (RMF) model, where the baryon-baryon interaction is usually mediated by the exchange of scalar ($\sigma$), vector ($\omega$) and isovector ($\rho$) mesons, and the hyperon-hyperon interaction could be incorporated through additional strange scalar ($\sigma*$) and vector ($\phi$) mesons. Hyperon coupling parameters in the model have been previously constrained by a variety of laboratory observables, and have the potential to be safely applied to high-density matter encountered in the stellar core. Studies based on several typical parameter sets have revealed no difficulties to allow a massive hyperon star as heavy as 2 $M_\odot$~\cite{Weissenborn12,Vid11}.

Especially, in one version of the RMF models, the $\sigma\omega\rho\phi$ model \cite{Weissenborn12}, the authors investigated consistently the vector
meson-hyperon coupling, going from the SU(6) quark model to a broader SU(3) symmetry. They varied the $g_8/g_1$ ratio and the vector meson interaction $\alpha_V$, with $g_1$ being the singlet coupling constant and $g_8$ being the octet coupling constant, in the following interaction Lagrangian between the meson nonet (singlet state $M_1$ and octet states $M_8$) and
the baryon octet $(B)$:
\begin{eqnarray}
   {\cal L}_{\it{int}}=&&
      - g_{8}\sqrt{2}
     \left[\alpha_V Tr \left( [\overline{B},M_8] B \right)
     +(1-\alpha_V) Tr \left( \{\overline{B},M_8 \} B \right) \right] \nonumber \\ &&- g_{1}{\sqrt{\frac{1}{3}}}
     Tr (\overline{B}B) Tr(M_1)
\end{eqnarray}

Hyperon star mass is generally believed to be smaller than the corresponding NS mass, because of the EoS softening of an extra degree of freedom. Most interestingly, what they found is a smooth transition between hyperon stars and NSs, as well as between their corresponding maximum masses. For example, for a certain value of parameter $\alpha_V$,  hyperons appear at successively higher densities with a smaller $g_8/g_1$ ratio, until $g_8/g_1=0$ the mass of a hyperon star comes as large as that of a NS. This motivates us to investigate systematically the fractional moment of inertia as a function of the hyperon coupling parameters.

Before we proceed, we mention that $\alpha_V$ is by definition restricted to the interval $0\sim1$, and $g_8/g_1$ is required to be $0\sim2/\sqrt{6}$, to be consistent with a experimentally-required repulsive vector $\omega$ coupling for all baryons . They are narrowed when the 2-solar-mass constraint is considered. We plot in Fig.~4 the fractional moments of inertia for different sequences of hyperon stars as a function of stars' mass, with $g_8/g_1=0,0.1,0.2,0.3$ with fixed $\alpha_V=1$, and $\alpha_V=0,0.3,0,6,0.9$ with fixed $g_8/g_1=1/\sqrt{6}$. We see that even when the crustal entrainment is chosen to be its lowest value ($0.016$), that is, neglecting completely its effect, one can derive useful limits for the hyperon coupling parameters from the condition $\bigtriangleup I/I\geq0.016$. In the case of $\alpha_V=1$, the choice of $g_8/g_1$ should not be smaller than $\sim 0.2$, that is a proper choice to be either $0.2$ or $0.3$. In the case of $g_8/g_1=1/\sqrt{6}$, $\alpha_V$ should not be smaller than $\sim 0.5$, that is a proper choice to be among $0.5 \sim 0.9$. These two limits ($\alpha_V=1, g_8/g_1\geq0.2$ and $g_8/g_1=1/\sqrt{6}, \alpha_V\geq0.5$) are really the lowest values possible since the $\bigtriangleup I/I$ value could be always larger than $0.016$, because of the presence of the crustal entrainment.

\section{Summary}

In this paper, we have confronted various most popular microscopic NS models with both the glitch data from Vela pulsar and the recent 2-solar-mass constraint from the mass measurements of PSR J1614-2230 and PSR J0348+0432. By relaxing the glitch crisis argument and taking the crustal entrainment as a free parameter between two critical cases of $\bigtriangleup I/I = 0.016$ (the case without entrainment) and $\bigtriangleup I/I =0.07$ (the case with entrainment), we aim to contribute a better understanding of the type of the Vela-like pulsars based on those microscopic models, and also select more reasonable hyperon coupling parameters for the widely-used phenomenological models, like the RMF model.

We choose two EoS models for the study of NSs and hybrid stars, respectively, guaranteeing each of them fulfilling the 2-solar-mass constraint. From the comparison of the results of these two types of NS models, we conclude that if the inferred $\bigtriangleup I/I$ value could be larger than the half of 0.07, namely a large enough fraction of neutrons are entrained, then it could not be a hybrid star. If the inferred $\bigtriangleup I/I$ value would be in the range of $0.016\sim 0.035$, as commonly estimated in the literature, a small fraction of free quarks could be possible in its core, but it will be difficult to determine it to be a NS or a hybrid star, because their predicted masses only differ $0.05\%$. Essentially this is due to the assumption that a hybrid star only has a small quark core and is composed mostly of nucleonic matter, as inferred by most of the model calculations with a high threshold of the deconfinement phase transition. The existence of another type of hybrid stars with a large quark core would complicate greatly this conclusion, and we refer to a future work to address carefully this point.

Then for the second objective of the present work, we switch to one of the widely-used phenomenological models, namely the $\sigma\omega\rho\phi$ RMF model. By only imposing the lowest possible value of the glitch constraint from Vela pulsar, namely $\bigtriangleup I/I\geq0.016$, one can effectively narrow the parameter space of the related hyperon coupling. Then a proper choice of the parameter sets can be as follows: $\alpha_V=1, g_8/g_1=0.2,0.3$ and $g_8/g_1=1/\sqrt{6}, 0.5\leq\alpha_V\leq0.9$. The results would be useful for subsequent studies on a hyperon star as an important type of NS model, to explain related pulsar observations based on this phenomenological model.

\section*{Acknowledgments}

\end{document}